\documentclass[aps,prl,twocolumn,groupedaddress,showpacs]{revtex4}
\usepackage{graphicx}
\usepackage{bm}

\begin{document}
\title{Experimental studies of the transient fluctuation 
theorem using liquid crystals}
\author{Soma Datta}
\author{Arun Roy}
\email{aroy@rri.res.in}
\affiliation{
   Raman Research Institute, C. V. Raman Avenue, Sadashivanagar,
   Bangalore 560 080, INDIA.
}
\date{\today}

\begin{abstract}
In a thermodynamical process, the dissipation or production of entropy
can only be positive or zero according to the second law of
thermodynamics. However the laws of thermodynamics are applicable to
large systems in the thermodynamic limit. Recently a fluctuation
theorem known as the Transient Fluctuation Theorem (TFT) which
generalizes the second law of thermodynamics even for small systems
has been proposed. This theorem has been tested in small systems such
as a colloidal particle in an optical trap.  We report for the first
time an analogous experimental study of TFT in a spatially extended
system using liquid crystals. 
\end{abstract}


\maketitle

The laws of thermodynamics describe the physical behaviour of
macroscopic systems. The second law of thermodynamics states that when
such a system is taken from one equilibrium state to another in a
process, the change in entropy can only be positive or zero depending
on wheather the process is irreversible or reversible
respectively. Though the laws of thermodynamics are applicable under
very general conditions, these laws are strictly valid only for large
systems in the so-called thermodynamic limit. For these large systems,
the effects of thermal noise on the average macroscopic physical
quantities are not manifested except under special physical conditions
such as near phase transitions.  However, when the system size is
small or more precisely the change in the relevant energy of the
system in a process is of the order of the thermal energy $k_BT$,
$k_B$ being the Boltzmann constant and $T$ being the absolute
temperature of the system, the thermal noise is expected to play an
important role on it's behaviour. In particular, the validity of the
second law of thermodynamics for small systems is of considerable
debate since the time of Boltzmann. Recently a nonequilibrium
fluctuation theorem (FT) known as the \emph{Transient Fluctuation
Theorem} (TFT) has been proposed to generalize the second law of
thermodynamics for these small systems\cite{evans93}.  In its most
general form, TFT not only predicts transient violation of second law
of thermodynamics when the dissipation is comparable to the thermal
energy $k_BT$ but also it provides an expression for the probability
that a dissipative flux flows in a direction opposite to that required
by the second law of thermodynamics. More precisely for a thermostated
system at temperature $T$, this theorem states that in a time interval
$\tau$, the probability $P(\Omega_\tau)$ of a dissipation
$\Omega_\tau$ being positive and the probability $P(-\Omega_\tau)$ of
same dissipation $\Omega_\tau$ being negative in an irreversible
process satisfies the following condition
\begin{equation}
\frac{P(\Omega_\tau)}{P(-\Omega_\tau)}=exp\left[ \frac{\Omega_\tau}{k_BT} \right].
\label{eq.1}
\end{equation}
The dissipation being an extensive quantity, the total dissipation
increases as either the system size or the observation time is
increased.  Then the above theorem implies that the production of
entropy or positive dissipation will be overwhelmingly more likely
than the consumption of entropy or negative dissipation in an
irreversible process for large systems or for large observation time
in accordance with the second law of thermodynamics. In this way the
TFT generalizes the second law of thermodynamics even for small
systems. Another FT for these small systems has also been proposed
which relates the work done along nonequilibrium trajectories, in
taking a system from one equilibrium state to another, to the
thermodynamic free energy difference between the initial and final
equilibrium states of the system\cite{jarzynski97}.  The physics of
such small systems has recently gained wide interest in the scientific
community\cite{retort03}. Examples of such system include
nano-materials, biological molecular machines, quantum dots etc. In
particular, the physics of such small systems when driven out of
equilibrium is of paramount importance in many technological
applications of these systems.

Though considerable theoretical and simulation studies\cite{gallavotti95,
kurchan98,narayan04,crooks00,williams04,seifert05} on these FTs have
been reported in the literature, only a few experimental studies on
the validity of these theorem have been performed. The main difficulty
in the experimental studies of these theorems is in arranging the
change in the relevant energy of the system under study to few $k_BT$
required for the system to show deviation from macroscopic laws of
thermodynamics.  Nevertheless, the validity of these fluctuation
theorems have been probed experimentally in small systems such as
colloidal particle in an optical trap\cite{wang02,carberry04}, in
two-level system\cite{schuler05}, torsion pendulum immersed in a
viscous fluid\cite{ciliberto06} and deformation dynamics of single RNA
molecule when stretched\cite{liphardt02}.  In fact these theorems have
been used to measure the folding free energy of a single RNA molecule
when stretched\cite{collin05} by an external applied force. However,
all these experiments are performed on relatively small systems. The
validity of these laws in the case of spatially extended systems where
the physical properties are described by an effective order parameter
which varies in space has not yet been probed experimentally. In this
letter, we report on the experimental studies of the TFT for such
spatially extended systems using nematic liquid crystals.

\begin{figure}
\includegraphics{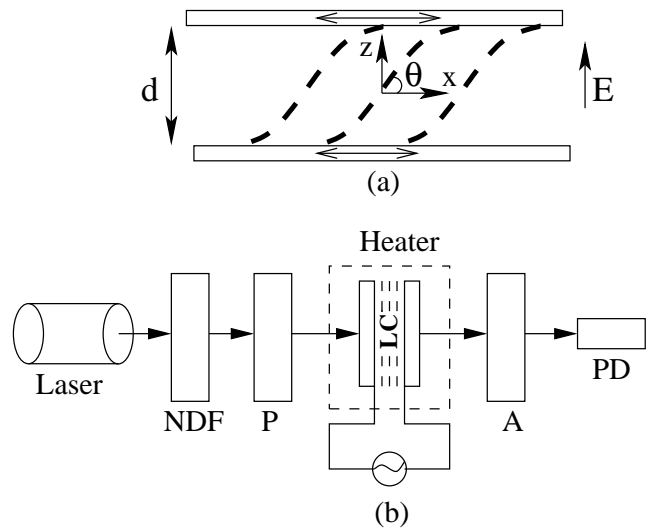}
\caption{Schematic representation of the experimental setup. (a) The
distorted director configuration in nematic liquid crystal between two
plates for applied electric field above Frederickz threshold. The
dashed lines represent the local orientation of $\bf n$ which makes an
angle $\theta$ with respect to the x-axis. The double arrow represents
the rubbing direction on the plates. (b) Schematic experimental setup
where the He-Ne laser beam passes through a neutral density filter
(NDF) and then through the LC sample held between crossed polarizers
(P) and (A).  The intensity of the light passing through the sample is
monitored using the photodiode (PD).}
\label{fig.1}
\end{figure}

Liquid crystals (LC) usually made of rod like organic molecules lack
the three-dimensional periodicity of crystals, but have anisotropic
physical properties\cite{degennes}. The simplest LC viz. Nematic LCs,
which are used in practically all commercial LC displays (LCD) have a
long range orientational order of the long axes of the molecules. The
orientational order described by the director $\bf n$ can be easily
deformed using external perturbations such as electric and magnetic
fields. The distortions free energy of the director field can be
described in terms of three elastic constants corresponding to the
splay, bend and twist types of distortions of the director
field\cite{degennes}.  In our experiments, a nematic liquid crystal is
sandwiched between two appropriately treated tin oxide coated glass
plates. The transparent tin oxide coating acts as electrode for the
application of an electric field without obstructing the optical
measurements. The glass plates are coated with polyimide and rubbed
along a certain direction. These treatments of the glass plates align
the long axes of the rod-like molecules parallel to the rubbing
direction giving rise to the so called homogeneous alignment of the
liquid crystals between the plates.  When such homogeneously aligned
nematic liquid crystals having positive dielectric anisotropy are
subjected to an applied electric field perpendicular to glass plates,
they exhibit a second order transition to a distorted structure
(fig.~\ref{fig.1}a) above a threshold electric field. This is known as
the Frederickz transition\cite{degennes}.
Above the Frederickz threshold, the equilibrium structure of the nematic
liquid crystal depends on the applied electric field.
  
As liquid crystals have anisotropic physical properties, the
refractive index $n_e$ for light polarized parallel to the director
$\bf n$ (extraordinary ray) is different from that of $n_o$
corresponding to light polarized perpendicular to $\bf n$ (ordinary
ray). This anisotropy in the refractive index of liquid crystals in
this geometry gives rise to a phase difference ($\Phi$) between the
extraordinary and ordinary ray of the light beam passing through the
sample and incident normal to the glass plates. The phase difference
$\Phi$ depends on the orientations of the director $\bf n$ between the
plates.  In the absence of the applied electric field, the director
$\bf n$ is parallel to the x-axis and $\Phi=2\pi\Delta nd/\lambda$ for
light propagating along the z-axis, where $\Delta n=n_e-n_o$ is the
birefringence of the LC sample and $d$, $\lambda$ are the thickness of
the LC sample and the wavelength of the light respectively. When the
applied electric field is increased beyond the Frederickz threshold,
the director $\bf n$ progressively becomes parallel to the electric
field and $\Phi$ decreases with increasing electric field.  Thus the
change in the director configuration can be monitored by measuring the
change in $\Phi$ using an interferometric technique under crossed
polarizers. We have experimentally studied the dissipative relaxation
dynamics of $\Phi$ when the liquid crystal is driven from one
equilibrium state to another by a step change in the applied electric
field. The change in the applied electric field is made sufficiently
small such that the free energy difference between the initial and
final state of the system is comparable to the thermal energy $k_BT$
and consequently study the validity of the TFT in this system.

The schematic experimental setup is shown in fig.~\ref{fig.1}b. The
homogeneously aligned liquid crystal cell is placed between crossed
polarizers with the director $\bf n$ making an angle 45 deg with
respect to the polarizer.  The intensity of a stabilized He-Ne laser
($\lambda=632.8 nm$) beam passing through the sample is monitored
using a high gain low noise photodiode.  The output of the photodiode
is digitized using a 16-bit data acquisition board at the rate of 1
kHz. We use a bipolar square wave electric field of frequency 1 kHz to
drive the sample from one equilibrium state to another by changing the
amplitude of the electric field. Use of AC electric field minimizes
the effects of the charge impurities invariably present in the liquid
crystals without affecting the director dynamics. The sample is
thermostated in a heater with temperature stability of $5 mK$.

\begin{figure}
\includegraphics[width=9cm]{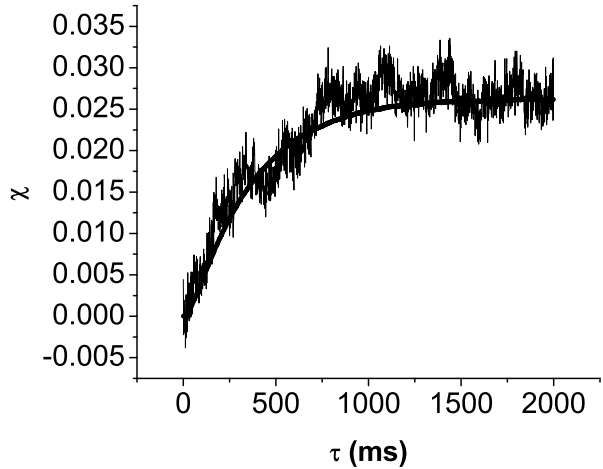}
\caption{The average and fluctuating relaxation dynamics of $\chi$
when the amplitude of the applied square wave voltage is changed from
$V_i=1.490 V$ to $V_f=1.488 V$. Averaging is done over 3000
experiments. The average dynamics is well described by a single
exponential decay to the final value with the relaxation time $362
ms$.}
\label{fig.2}
\end{figure}

The experiments are performed on nematic liquid crystal (5CB) for
sample thickness $d=27.0 \mu m$. The experiments are carried out at
$30^\circ C$ which is about $5^\circ C$ below the transition
temperature ($35.1^\circ C$) from the isotropic to the nematic
phase. For a given temperature, we hold the sample with the amplitude
of the applied square wave voltage at a certain value ($V_i$) for 2
seconds, then the amplitude of the applied voltage is changed in a
step to a slightly lower value ($V_f$) and held at that value for
subsequent 2 seconds. The normalized phase difference
$\chi=(\Phi-\bar{\Phi}_i)/\bar{\Phi}_i$, where $\bar{\Phi}_i$ is the
equilibrium value of $\Phi$ at the initial voltage $V_i$, is monitored
as it relaxes from it's initial equilibrium value of zero to the final
equilibrium value $\bar{\chi}_f$ at voltage $V_f$. This cycle is
repeated for 3000 times to find the probability distributions of the
dissipation over an ensemble of experiments.  Fig.~\ref{fig.2} shows
the average relaxation dynamics of $\chi$ averaged over 3000
experiments and the fluctuating dynamics of $\chi$ for one of the
experiments when the amplitude of the applied voltage is changed from
$V_i=1.490 V$ to $V_f=1.488 V$.
The average dynamics of $\chi$ is described by a single exponential
decay from the initial value to the final value with a characteristic
relaxation time $\tau_r\sim 362 ms$.  Thus the dynamical evolution of
$\chi$ is analogous to the relaxation dynamics of a particle in a
harmonic potential in the overdamped regime in the presence of thermal
noise described by Langevin equation. The effective strength
($k_\chi$) of the harmonic potential associated with $\chi$ for a
given applied voltage can be determined by measuring the thermal
fluctuation of $\chi$ and by the application of the equipartition
theorem $k_\chi=k_BT/\sigma^2_\chi$ where $\sigma^2_\chi$ is the
variance of $\chi$. For this purpose, we sample $\chi$ in the initial
and final equilibrium states of the system for 40 seconds and
determine $k_\chi$ from the measured values of the variance
$\sigma^2_\chi$ in these states.  As $V_i-V_f=0.002 V$ is small in our
experiments, the measured values of $\sigma^2_\chi$ and hence that of
$k_\chi$ in the initial and final state of the system are found to be
equal within the experimental error. The measured value of the
variance $\sigma^2_\chi=1.13\times 10^{-7}$ for $1.490 V$ at $30
^\circ C$.

\begin{figure}
\includegraphics[width=9cm]{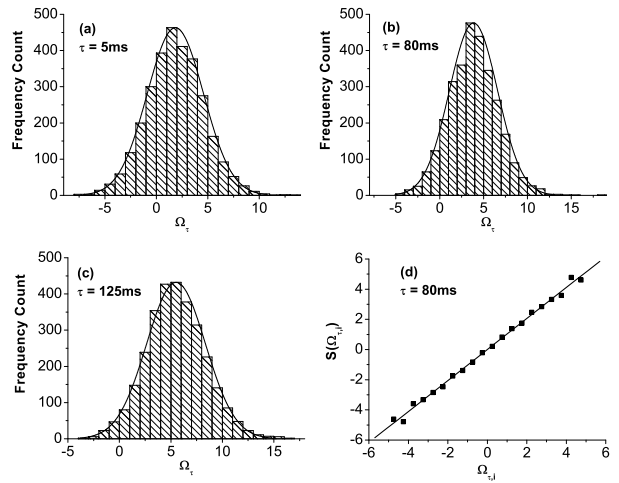}
\caption{Histograms of the dissipation $\Omega_\tau$ obtained from an
ensemble of 3000 experiments at different $\tau$ (a) $5 ms$, (b) $80
ms$ and (c) $125 ms$.  The bin size $\Delta=1.0$. The solid lines are
the fit to the normal distributions.  (d) The corresponding plot of
$S(\Omega_{\tau i})$ versus the the dissipation $\Omega_{\tau i}$
associated with the i-th histogram bin for $\tau=80 ms$. The solid
line is the best linear fit with slope $1.02$ which agrees very well
with the TFT predicted value of one.}
\label{fig.3}
\end{figure}

The evolution of $\chi$ after the step-like change in the applied
voltage in this system is analogous to the overdamped Langevin type
dynamics of a particle from the initial equilibrium position
zero to final equilibrium position $\bar{\chi}_f$, when the harmonic 
potential felt by the particle is shifted from position zero to 
$\bar{\chi}_f$, keeping the strength of the potential $k_\chi$ same.
Using the above analogy, the dissipation in time $\tau$ can be 
written as\cite{carberry04,reid04}
\begin{equation}
\Omega_\tau = \frac{k_\chi \bar{\chi}_f}{k_BT}(\chi_\tau-\chi_0).
\label{eq.2}
\end{equation}
where $\chi_0$ and $\chi_\tau$ are the values of $\chi$ at time zero
and $\tau$ respectively.  As expected, the expression for the
dissipation $\Omega_\tau$ is same as the work done in
time $\tau$ by a constant force $k_\chi \bar{\chi}_f$ on a particle
which is in a harmonic potential of strength 
$k_\chi$\cite{kurchan98,narayan04}.  We calculate the dissipation
$\Omega_\tau$ from Eq.~(\ref{eq.2}) for the 3000 repetitions of the
experiment to get the desired histogram of $\Omega_\tau$.  The
histograms of $\Omega_\tau$ for $\tau=5ms$,$80 ms$ and $125 ms$ are
shown in fig.~\ref{fig.3}.
For $\tau$ small, the average dissipation is small. Therefore the
probability distribution of $\Omega_\tau$ is peaked near zero with
considerable probability of observing negative dissipation as can be
seen from fig.~\ref{fig.3}a.  As $\tau$ increases, the average
dissipation increases and the peak of the probability distribution of
$\Omega_\tau$ shifts toward positive values of $\Omega_\tau$
(fig.~\ref{fig.3}b, fig.~\ref{fig.3}c) and the probability of
observing negative dissipation decreases.  For very large $\tau$, the
dissipation is much larger than $k_BT$ and the probability of
observing negative dissipation becomes negligible as required by
second law of thermodynamics.

\begin{figure}
\includegraphics[width=9cm]{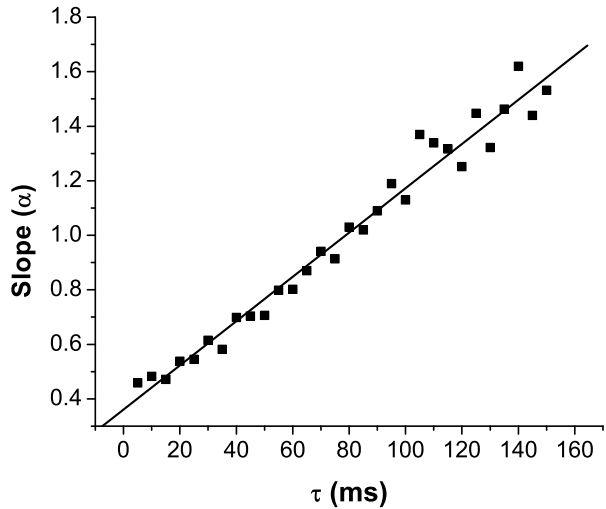}
\caption{The variation of the slope $\alpha$ with observation time
$\tau$. $\alpha$ is close to the TFT predicted value of one for $\tau$
close to $80 ms$. The solid line is the best linear fit described by
the equation $\alpha= 8.12 \times 10^{-3}\tau (ms) + 0.36$.}
\label{fig.4}
\end{figure}

To test the validity of the TFT in this system, we evaluate the
function $S(\Omega_\tau)=ln[P(\Omega_\tau)/P(-\Omega_\tau)]$ from the
experimentally measured probability distribution function of
$\Omega_\tau$. Then according to the TFT, $S(\Omega_\tau)$ should be a
linear function of $\Omega_\tau$ i.e.
$S(\Omega_\tau)=\alpha\Omega_\tau$ with slope $\alpha=1.0$ for all
$\tau$.  We find from the experimentally measured histogram of
$\Omega_\tau$ the probability count ($N_i$) for dissipation between
$\Omega_{\tau i}-\Delta/2$ and $\Omega_{\tau i}+\Delta/2$
corresponding to the i-th histogram bin $\Omega_{\tau i}=i\Delta$,
$\Delta$ being the bin size.  Then the experimental analog of the
function $S(\Omega_{\tau i})=ln(N_i/N_{-i})$.  We find that the
experimentally determined $S(\Omega_{\tau i})$ can be fitted well by a
linear function of $\Omega_{\tau i}$ for all $\tau$.  In
fig.~\ref{fig.3}d, we show the plot of $S(\Omega_{\tau i})$ verses the
dissipation $\Omega_{\tau i}$ for $\tau=80 ms$.  The straight line in
fig.~\ref{fig.3}d shows the corresponding linear fit to the data with
slope $\alpha=1.02$ agreeing very well with the prediction of
TFT. Though the experimental results agree very well with the
prediction of TFT for $\tau$ close to $80 ms$, we find that in this
system the value of the slope $\alpha$ is not equal to one for all
$\tau$ but it increases linearly with $\tau$. Fig.~\ref{fig.4} shows
the variation of the slope $\alpha$ with $\tau$ which can be fitted
with a linear equation $\alpha= 8.12 \times 10^{-3}\tau (ms) + 0.36$.
We have repeated the experiments at different temperatures and for
different applied voltages and we have observed similar variation of
$\alpha$ with $\tau$. The details of these experimental results will
be published elsewhere.

In conclusion we have experimentally studied the validity of the TFT
in the spatially extended system using liquid crystals.  The TFT is
studied for the dissipative dynamics of a macroscopic order parameter
\emph{viz.} the orientational order of the liquid crystals when driven
from one equilibrium state to another.  We construct the appropriate
dissipation function for this system and experimentally determine the
histograms of the dissipation over an ensemble of experiments. The
experimental results are compared with the predictions of the TFT. We
find agreement between the experimental observations and the
predictions of TFT only for particular values of the observation time
$\tau$. Simulation studies of spatially extended systems may shed
further insight on to the validity of TFT in these systems.  We thank
Dr. Abhishek Dhar for some helpful discussions and comments on the
subject.

\end{document}